\documentclass{aa}

\def\puncspace{\ifmmode\,\else{\ifcat.\C{\if.\C\else\if,\C\else\if?\C\else%
\if:\C\else\if;\C\else\if-\C\else\if)\C\else\if/\C\else\if]\C\else\if'\C%
\else\space\fi\fi\fi\fi\fi\fi\fi\fi\fi\fi}%
\else\if\empty\C\else\if\space\C\else\space\fi\fi\fi}\fi}
\def\SP{\let\\=\empty\futurelet\C\puncspace}

\def\h1{$h^{-1}$\SP}

\def\etal{{\it et al.\/}\ }

\def\eg{{\it e.g.\/}\rm,\ }
\def\lsim{~\rlap{$<$}{\lower 1.0ex\hbox{$\sim$}}}
\def\gsim{~\rlap{$>$}{\lower 1.0ex\hbox{$\sim$}}}
\def\void#1{{}}

\usepackage{graphics}

\begin{document}

   \thesaurus{(11.03.1); (12.12.1); (12.03.3)}     

   \title{ESO Imaging Survey}

   \subtitle{VII. Distant Cluster Candidates over 12 square degrees}

\author { M. Scodeggio\inst{1} \and L.F. Olsen\inst{1,2} 
 \and L. da Costa\inst{1} \and R. Slijkhuis\inst{1,3} 
\and C. Benoist\inst{1}
\and E. Deul\inst{1,3} \and T. Erben\inst{1,4}  \and R. Hook\inst{5} \and M. Nonino \inst{1,6}  \and A. Wicenec\inst{1} \and S. Zaggia \inst{1,7}}

\offprints{M. Scodeggio}

\institute{
European Southern Observatory, Karl-Schwarzschild-Str. 2,
D--85748 Garching b. M\"unchen, Germany \and
Astronomisk Observatorium, Juliane Maries Vej 30, DK-2100 Copenhagen, 
Denmark \and
Leiden Observatory, P.O. Box 9513, 2300 RA Leiden, The Netherlands \and
Max-Planck Institut f\"ur Astrophysik, Postfach 1523 D-85748,  Garching b. 
M\"unchen, Germany \and
Space Telescope -- European Coordinating Facility, Karl-Schwarzschild-Str. 
2, D--85748 Garching b. M\"unchen, Germany \and
Osservatorio Astronomico di Trieste, Via G.B. Tiepolo 11, I-31144
Trieste, Italy \and
Osservatorio Astronomico di Capodimonte, via Moiariello 15, I-80131 Napoli,
Italy
}

\date{Received ; accepted }

\maketitle

\begin{abstract}

In this paper the list of candidate clusters identified from the
I-band data of the ESO Imaging Survey (EIS) is completed using the
images obtained over a total area of about 12 square degrees.
Together with the data reported earlier the total I-band coverage of
EIS is 17 square degrees, which has yielded a sample of 252 cluster
candidates in the redshift range $0.2 \lsim z \lsim 1.3$. This is the largest
optically-selected sample currently available in the Southern
Hemisphere. It is also well distributed in the sky thus providing
targets for a variety of VLT programs nearly year round.

\end{abstract}


\section{Introduction}
\label{sec:introduction}

The discovery of clusters of galaxies at high redshifts has motivated
efforts of compiling lists of candidates for follow-up observations
with 8m-class telescopes. The interest in studying these systems spans
a broad range of topics and searching for them was identified as one
of the primary goals of the ESO Imaging Survey (EIS, Renzini \& da
Costa 1997), a moderately deep wide-field imaging survey recently
conducted at the 3.5m-NTT telescope at La Silla. The main requirements
for the cluster search were: 1) to produce a list of candidates large
enough to meet the needs of potential VLT programs; 2) to span a broad
range of redshifts; 3) to cover a wide range of right ascension
thereby allowing the selection of targets year round; 4) to minimize
as much as possible spurious detections. These requirements dictated
to a large extend the observing strategy adopted by EIS such as the
selection of four fields and the preference given to I-band
observations in the second-half of the program. While searches in
other wavelengths may provide less contaminated and better defined
samples (\eg IR and X-ray searches), optical searches have the
advantage of producing large samples at a faster rate than any other
search method, especially with the advent of CCD wide-field imagers.

As stated in Olsen \etal (1998a; paper~II) the main goal of the EIS
cluster search program is to {\it timely} provide the astronomical
community with a list of cluster candidates that can be used as
individual targets for follow-up observations in the Southern
Hemisphere, especially with the VLT. It must be emphasized that it is
not the intention of the present paper to provide a complete and
well-defined sample for statistical studies, since such analysis is
beyond the scope of the present effort.

The original aim of EIS was to observe about 20 square degrees in four
different patches of the sky as described by Renzini and da Costa
(1997) (see also Nonino \etal 1998, paper~I), with a significant
fraction of the area covered in V and I-bands.  However, as described
in earlier papers (paper~I, Prandoni \etal 1998; paper~III) the
first-half of the program was severely compromised by bad
weather. Therefore, in the second-half, I-band observations covering
EIS patches C and D were given priority. The data for these patches
are far superior than those of earlier patches and the full coverage
of the pre-selected areas was possible, yielding a total area of about
12 square degrees (Benoist \etal 1998, paper~VI). In this paper the
list of cluster candidates found in these regions by using the cluster
finding pipeline described in paper~II is presented. The results
extend the candidate cluster sample presented in papers~II and Olsen
\etal (1998b, paper~V), providing targets nearly year round.

In section 2 some aspects of the data relevant to the application of
the cluster detection algorithm are discussed. In section 3 a list of
198 candidate clusters is presented and the results compared with
those of other patches and of the Palomar Distant Cluster Survey
(PDCS, Postman \etal 1996), currently the only comparable survey. A
brief summary is presented in section 4.

\section {Galaxy Catalogs}

The generation and the characteristics of the EIS galaxy catalogs in
patches C and D have been discussed in paper~VI. In that paper they
were shown to be considerably more homogeneous than those derived from
previous patches, with only small variations in depth. As in previous
papers, the odd and even catalogs extracted from single exposure
images (paper~I and II) were independently used to identify clusters
of galaxies. This was done by applying the matched filter algorithm
described in paper~II to six overlapping sections of approximately the
same size covering each of the patches considered.  For patch C the
sections were chosen to avoid a small ($\sim0.2$ square degree)
shallow region mentioned in paper~VI. In order to guarantee a full
overlap between the regions covered by the odd and even frames the
edges of the patches were also trimmed yielding an effective are of
5.3 and 5.5 square degrees for patches C and D, respectively.

The first set of candidate clusters derived from the even and odd
frames consisted of over 100 objects in each patch. However, these
included an unusually large number of unpaired highly significant
detections. The visual inspection of all candidate clusters, together
with the even and odd galaxy catalogs, showed that the observed
asymmetries were due to the presence of spurious objects detected in
the vicinity of bright, saturated stars. As pointed out in paper VI,
the reason for this is possibly an electronic problem of the old EMMI
controller, when used in the dual-port readout mode. This problem
affected the last three runs of EIS by producing faint light
trails associated with saturated stars, when these are located in the
lower-half part of the detector. Along the trail a large number of
spurious, low surface brightness objects are identified, affecting
either the odd or even frames in different parts of the sky, but not
both simultaneously for the same star. The fraction of these objects
is relatively small and they do not significantly affect the number
counts or correlation function. However, they have a significant
impact in the performance of the matched-filter algorithm which
identifies a large number of cluster candidates near bright
stars. Since patches C and D are located at low galactic latitudes,
with a large number of saturated stars, the frequency of the problem
is large, affecting about 50\% of the original detections.

Fortunately, the above problem can be partially overcome taking
advantage of the sampling strategy of the survey whereby each position
on the sky is sampled at least twice by different parts of the
detector (paper~I). Since from these single exposures two catalogs are
derived, it is possible to overcome the light-trail problem at the
catalog level by using, instead of the odd and even catalogs, the
catalog which only includes paired galaxies (hereafter, referred to as
the paired catalog), detected in both of these catalogs. By
construction, this eliminates most spurious objects. The only
disadvantage with this procedure is that only one catalog of candidate
clusters can be produced and the galaxy sample is slightly
shallower. Of course, this solution cannot be applied to samples
extracted from the co-added images, which will therefore require some
type of correction at the image level.  Various alternatives are
currently being considered.

\section {Catalog of Cluster Candidates}

The paired catalogs for the two patches were produced and used as
input to the cluster finding pipeline using the same sections and
parameters as in the case of the odd/even catalogs. As expected, the
use of paired catalogs avoids all cases of cluster candidates that had
been detected in the vicinity of light trails and occasionally faint
satellite tracks. Note that new candidates are also found, probably
because of subtle changes in the background population. It is worth
emphasizing that visual inspection of these new candidates shows that
they are in general very robust. In order to take advantage of these
new detections the final cluster candidate list, shown below, is a
combination of all $\geq 3\sigma$ detections derived from the odd/even
catalogs, eliminating the spurious detections described in the
previous section, supplemented by those derived from the paired
catalog.

\begin{figure*}
\resizebox{\textwidth}{!}{\includegraphics{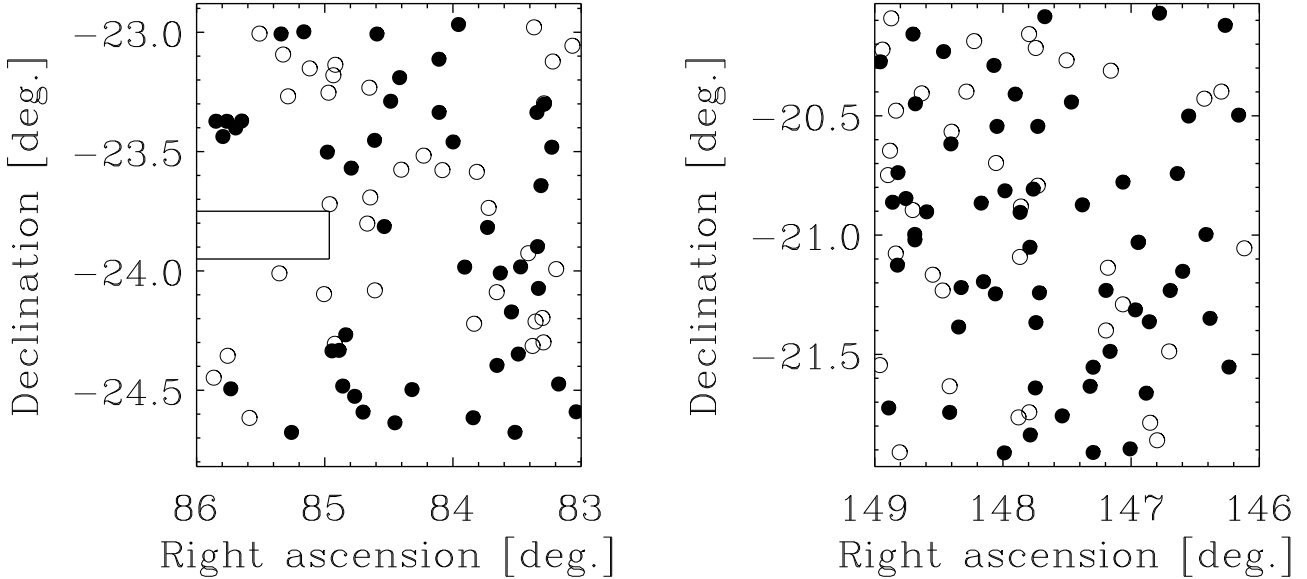}}
\caption{The projected distributions for the cluster
candidates detected in Patches~C (left panel) and D (right
panel). The filled circles mark the distributions for the ``good''
candidates as defined in the text. In the distribution for patch C the
region discarded from the analysis is indicated.}
\label{fig:proj_dist}
\end{figure*}

Table~\ref{tab:cluster1} lists 103 cluster candidates in patches~C and
D detected either at 4$\sigma$ in one or at $3 \sigma$ in both
odd/even catalogs. These were the objects considered as ``good''
candidates in papers II and V.  Note that 69\% of them were also
detected using the paired catalog. Table~\ref{tab:cluster2} lists the
56 candidates which were detected at $3\sigma$ in only one of the
even/odd catalogs and in some cases at lower significance in the
other. In addition, in contrast to the previous papers, the table also
includes 39 candidates, corresponding to 20\% of the total sample,
which were only detected in the paired catalog. The tables give: in
column (1) the identification; in columns (2) and (3) the right
ascension and declination in J2000; in column (4) the estimated
redshift; in columns (5) and (6) two measures of the cluster richness
(see paper~II); in columns (7) and (8) the significance of the
detection in the even and odd catalogs, respectively; and finally in
column (9) the significance of the detection in the paired catalog. In
the case of high-z clusters $m_3 + 2$ might exceed the limiting
magnitude of the catalog, and no estimate of $N_R$ is possible. These
cases are indicated by $N_R = -99$.

In paper~II the frequency of noise peaks was estimated to be 0.4 per
square degree for the $4\sigma$ detections and 4.6 per square degree
for the $3\sigma$ detections. Therefore the contamination by spurious
detections of the total sample presented in tables~\ref{tab:cluster1}
and \ref{tab:cluster2} is expected to be $\sim$25\% with a
significantly smaller frequency if only table~\ref{tab:cluster1} is
considered.

All detections have been visually inspected and nearly all appear to
be promising candidates, although the reliability of the low-redshift
candidates is usually more difficult to evaluate.  As pointed out
above, candidates detected in the paired catalog are particularly
encouraging.  Furthermore, high-redshift clusters are more frequent in
the paired catalog than in the odd/even catalogs. This is probably
because the galaxy pairing eliminates faint spurious objects. It
should be pointed out that there are also cases where a cluster is
detected in either one or both odd/even catalogs but is not detected
in the paired catalog. This is possibly due to more subtle effects in
the background and noise properties of the maximum likelihood maps.
In other cases, especially those detected at relatively high
significance in one set but not in the other, the center of the
candidate cluster and/or the redshift estimate appear to be
incorrect. This is most likely due to projection effects of clusters
lying along the line-of-sight, which are not well resolved by the
searching algorithm. Finally, note that in patches C and D about 90\%
of the ``good'' candidates are detected in both the even and odd
catalogs, in contrast to the 65\% found in patches A and B.  This
better matching of detections is because the data for patches C and D
are significantly more homogeneous than those of previous patches.


Out of the 198 candidates listed in tables~\ref{tab:cluster1} and
\ref{tab:cluster2}, 93 are in patch~C and 105 in patch~D, over an
effective area of 5.3 and 5.5 square degrees, respectively. The
implied number density of clusters is about 18.3 clusters/square
degree slightly higher than the values found for patch~A and by
Postman \etal (1996), but similar to the value found for patch~B.

The projected distributions of the cluster candidates over the two
patches are shown in figure~\ref{fig:proj_dist}. As can be seen in
this figure the candidates appear to be distributed uniformly over the
whole area of the patches, independently of their significance.  Also
note that in patch C there are two groupings of cluster candidates, one
at $\alpha=05^h 43^m 30^s$,$\delta = -23^\circ 25'$ and the other at
$\alpha=05^h 40^m 00^s$,$\delta = -24^\circ 18'$. In the first case
there are three detections at $z \sim 0.2$ and two at $z \sim
0.6-0.7$. If confirmed the lower redshift system would be an
interesting region for follow-up work. The second system corresponds
to an overlap of four systems roughly along the same line-of-sight
but with redshifts ranging from 0.2 to 0.6.  These systems correspond
to the cases with large uncertainties in the position of the center
and redshift estimate observed during the visual inspection, as
discussed in the previous section.

\begin{figure}
\resizebox{\columnwidth}{!}{\includegraphics{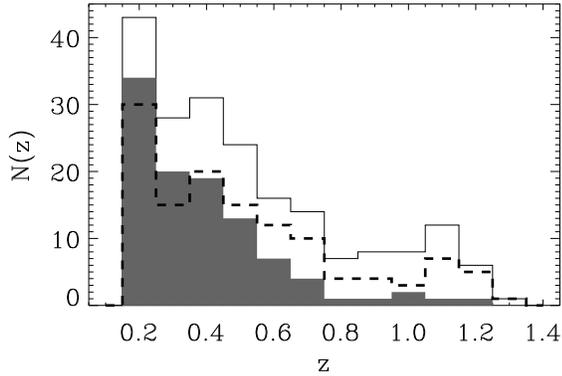}}
\caption{The redshift distribution for the cluster candidates
detected in Patches~C and D. The shaded area marks the
distribution for the ``good'' candidates as defined in the text. The
dashed line shows the distribution for the candidates detected in the
paired catalogs.}
\label{fig:z}
\end{figure}

\begin{figure}
\resizebox{\columnwidth}{!}{\includegraphics{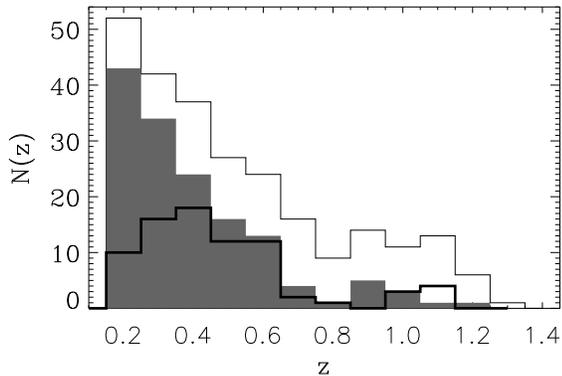}}
\caption{The redshift distribution of the total sample of EIS clusters
(upper panel) as presented in the present work and in Papers~II and V
in total covering an area of $\sim 14.4$ square degrees. The thick
line shows the estimated redshift distribution of cluster candidates
in the PDCS, covering 5.1 square degrees.}
\label{fig:zall}
\end{figure}

Figure~\ref{fig:z} shows the estimated redshift distribution of the
combined sample of candidate clusters identified in patches~C and
D. The median redshift for this sample is 0.4, which is comparable to
the values found for patches~A and B. Although a similar median
redshift is found for the candidates obtained from the paired catalog,
their redshift distribution (shown in the figure as the dashed line)
shows a tail at the high-redshift end ($z \gsim 0.8$) in contrast to
what is seen for the redshift distribution of the good candidates
found in the even/odd catalogs (indicated by the shaded area).  Recall
that the intrinsic uncertainty of the estimated redshifts is no less
than 0.1, due to the discreteness of the filter redshift values
(paper~II). Furthermore, because of the minimal overlap with clusters
with known redshift, the absolute accuracy of the redshift estimates,
produced by the cluster finding pipeline, cannot be easily
quantified. Therefore the current redshift estimates should be
considered tentative, until spectroscopic observations become
available.

The total sample of EIS cluster candidates, obtained by combining the
detections in the four EIS-wide patches, consists of 252 objects in
the redshift range $0.2 \leq z \leq 1.3$. The redshift distribution of the
combined sample is shown in figure~\ref{fig:zall}. The median redshift
of the distribution is $z\sim 0.4$. Note that the EIS redshift
distribution differs somewhat from that observed by PDCS, also shown
in the figure. The number of EIS candidates decreases monotonically
with redshift up to $z \sim 0.6$ with an extended tail beyond, in
contrast to the PDCS which shows a relatively flat distribution
peaking at $z \sim 0.4$.

\section{Summary}

This paper completes the presentation of one of the primary products
of EIS, namely a large sample of candidate clusters of galaxies
spanning a broad range of redshifts, extending to $z \sim 1$.  The
candidates were selected in four different patches of the sky covering
a large range in right ascension, thereby providing potential targets
for VLT which are observable over almost the entire year. Taking all
patches together the total sample consists of 252 candidates with
about 100 candidates with $z \gsim 0.5$. This is by far the largest
such a sample currently available and should serve as a good starting
point for several programs at the VLT. Note that as emphasized in
previous papers of this series the selection criteria adopted has been
in general conservative and the primary concern has been the
reliability of the candidates rather than completeness of the sample.
The catalogs of cluster candidates are available at
"http:://www.eso.org/eis" from where image cutouts
from the co-added image can be retrieved for evaluation and
preparation of follow-up observations.

Recall that the current cluster candidate lists have been prepared
based on galaxy catalogs extracted from the single 150 sec exposures.
Since these images are being co-added in the near future it will be
possible to extract galaxy catalogs which should reach about 0.5 mag
deeper.  As soon as these catalogs become available they will also be
used to search for clusters and it might be possible to extend
somewhat the redshift range of the detected cluster candidates and/or
confirm previous detections.  However, the available sample is
sufficiently large and deep to meet most of the scientific needs in
the first year of operation of the VLT.

\void{
This is by far the largest such a sample currently available and will
certainly be of great value for future observations in the southern
hemisphere. It is also spread over a wide range of right ascension
thus providing targets observable almost year round.
}

\begin{acknowledgements}
 
The data presented here were taken at the New Technology Telescope at
the La Silla Observatory under the program IDs 59.A-9005(A) and
60.A-9005(A). We thank all the people directly or indirectly involved
in the ESO Imaging Survey effort. In particular, all the members of
the EIS Working Group for the innumerable suggestions and constructive
criticisms. We also thank the ESO Archive Group and ST-ECF for their
support. Our special thanks to A. Renzini, VLT Programme Scientist,
for his scientific input, support and dedication in making this
project a success. Finally, we would like to thank ESO's Director
General Riccardo Giacconi for making this effort possible.
 
\end{acknowledgements}

\begin{table*}
\caption{The $4\sigma$ or paired cluster candidates for EIS patches C
and D.}
\begin{tabular}{lr@{\extracolsep{1mm}}r@{\extracolsep{1mm}}rr@{\extracolsep{1mm}}r@{\extracolsep{1mm}}rrrrrrr}
\hline \hline
Cluster name & \multicolumn{3}{c}{$\alpha$ (J2000)} &
\multicolumn{3}{c}{$\delta$ (J2000)} & $z$ & $\Lambda_{cl}$ & $ N_R $ & $\sigma_{even}$ & $\sigma_{odd}$ & $\sigma_{paired}$ \\ 
\hline
EIS 0532$-$2435 & 05 & 32 & 21.7 & $-$24 & 35 & 25.3 & 0.4 &  49.2 &   23 &  3.3 &  4.1&  4.2\\ 
EIS 0532$-$2428 & 05 & 32 & 55.6 & $-$24 & 28 & 26.8 & 0.5 &  42.2 &   36 &  4.2 &  2.8 & 3.5\\ 
EIS 0533$-$2328 & 05 & 33 &  8.9 & $-$23 & 28 & 54.0 & 0.4 &  60.9 &   62 &  5.3 &  5.5 & 5.6\\ 
EIS 0533$-$2318 & 05 & 33 & 23.5 & $-$23 & 18 &  3.8 & 0.4 &  35.8 &   37 &  3.9 &  3.7 & $-$\\ 
EIS 0533$-$2338 & 05 & 33 & 29.9 & $-$23 & 38 & 33.2 & 0.4 &  49.6 &   45 &  4.5 &  4.4 & 5.2\\ 
EIS 0533$-$2404 & 05 & 33 & 34.7 & $-$24 &  4 & 24.1 & 0.4 &  57.9 &   59 &  4.9 &  $-$ &  $-$\\ 
EIS 0533$-$2353 & 05 & 33 & 36.5 & $-$23 & 53 & 52.9 & 0.7 &  72.0 &   21 &  3.1 &  3.3 & 3.6\\ 
EIS 0533$-$2320 & 05 & 33 & 37.6 & $-$23 & 20 &  9.5 & 0.3 &  36.7 &   38 &  4.4 &  4.0&  4.7\\ 
EIS 0534$-$2358 & 05 & 34 &  9.2 & $-$23 & 58 & 59.6 & 0.6 &  71.6 &   22 &  3.6 &  3.9&  4.2\\ 
EIS 0534$-$2420 & 05 & 34 & 13.7 & $-$24 & 20 & 54.5 & 0.2 &  42.1 &   10 &  6.0 &  6.6 & $-$\\ 
EIS 0534$-$2440 & 05 & 34 & 20.1 & $-$24 & 40 & 35.5 & 0.4 &  54.1 &   54 &  4.5 &  4.5 & 4.6\\ 
EIS 0534$-$2410 & 05 & 34 & 26.9 & $-$24 & 10 & 19.1 & 0.3 &  40.2 &   40 &  3.6 &  5.5 & 4.6\\ 
EIS 0534$-$2400 & 05 & 34 & 48.5 & $-$24 &  0 & 32.4 & 0.2 &  31.0 &   29 &  5.2 &  5.0 &  5.9\\ 
EIS 0534$-$2423 & 05 & 34 & 55.0 & $-$24 & 23 & 45.5 & 0.2 &  41.4 &   26 &  7.9 &  6.4 & 6.9\\ 
EIS 0535$-$2349 & 05 & 35 & 13.2 & $-$23 & 49 &  4.2 & 0.4 &  49.1 &   30 &  4.3 &  3.7 & 4.6\\ 
EIS 0535$-$2436 & 05 & 35 & 41.1 & $-$24 & 36 & 55.1 & 0.3 &  35.8 &   23 &  3.9 &  4.2 & $-$\\ 
EIS 0535$-$2359 & 05 & 35 & 57.4 & $-$23 & 59 &  2.7 & 0.8 & 151.5 &   31 &  3.7 &  4.3 & $-$\\ 
EIS 0536$-$2258 & 05 & 36 &  9.4 & $-$22 & 58 &  3.2 & 1.1 & 210.8 &  $-$99 &  3.8 &  3.0 & 3.1\\ 
EIS 0536$-$2327 & 05 & 36 & 19.8 & $-$23 & 27 & 34.5 & 0.3 &  54.4 &   26 &  8.2 &  5.1&  7.7\\ 
EIS 0536$-$2320 & 05 & 36 & 46.9 & $-$23 & 20 &  8.8 & 0.4 &  34.7 &   21 &  3.7 &  3.0&  $-$\\ 
EIS 0536$-$2306 & 05 & 36 & 47.1 & $-$23 &  6 & 46.6 & 0.2 &  19.7 &   23 &  5.2 &  5.4 &  5.8\\ 
EIS 0537$-$2429 & 05 & 37 & 39.4 & $-$24 & 29 & 50.1 & 0.5 &  43.0 &   40 &  3.2 &  3.2 & $-$\\ 
EIS 0538$-$2311 & 05 & 38 &  3.2 & $-$23 & 11 & 24.2 & 0.6 &  99.3 &   87 &  6.9 &  6.7 & 7.3\\ 
EIS 0538$-$2438 & 05 & 38 & 12.4 & $-$24 & 38 & 11.4 & 0.3 &  38.1 &   27 &  4.4 &  4.6 & 5.1\\ 
EIS 0538$-$2317 & 05 & 38 & 20.6 & $-$23 & 17 & 20.7 & 0.7 &  81.7 &  109 &  2.9 &  4.2&  $-$\\ 
EIS 0538$-$2348 & 05 & 38 & 33.0 & $-$23 & 48 & 50.5 & 0.7 &  74.6 &   83 &  4.2 &  3.4 & 3.1\\ 
EIS 0538$-$2300 & 05 & 38 & 47.1 & $-$23 &  0 & 26.4 & 0.5 &  44.4 &   19 &  3.0 &  3.5 &  3.9\\ 
EIS 0538$-$2327 & 05 & 38 & 51.7 & $-$23 & 27 & 10.4 & 0.2 &  26.8 &   19 &  5.8 &  8.0 & $-$\\ 
EIS 0539$-$2435 & 05 & 39 & 14.0 & $-$24 & 35 & 30.0 & 0.2 &  32.8 &   19 &  4.6 &  5.1 & 5.0\\ 
EIS 0539$-$2431 & 05 & 39 & 30.3 & $-$24 & 31 & 32.2 & 0.4 &  49.9 &   54 &  4.5 &  4.4&  4.6\\ 
EIS 0539$-$2334 & 05 & 39 & 36.9 & $-$23 & 34 &  9.3 & 0.2 &  26.7 &   23 &  5.8 &  5.1 & $-$\\ 
EIS 0539$-$2416 & 05 & 39 & 47.5 & $-$24 & 16 &  3.8 & 0.3 &  31.5 &   17 &  3.9 &  3.0 & $-$\\ 
EIS 0539$-$2428 & 05 & 39 & 53.3 & $-$24 & 28 & 57.4 & 0.6 &  71.8 &   24 &  3.7 &  3.8 & 4.3\\ 
EIS 0540$-$2419 & 05 & 40 &  0.3 & $-$24 & 19 & 58.6 & 0.2 &  33.7 &   22 &  5.0 &  5.4 & 6.1\\ 
EIS 0540$-$2420 & 05 & 40 & 14.0 & $-$24 & 20 &  6.1 & 0.4 &  40.3 &   11 &  $-$ &  4.2 & $-$\\ 
EIS 0540$-$2330 & 05 & 40 & 23.1 & $-$23 & 30 &  8.1 & 0.3 &  31.5 &   41 &  4.1 &  3.4 & $-$\\ 
EIS 0541$-$2259 & 05 & 41 &  8.6 & $-$22 & 59 & 53.0 & 0.2 &  37.5 &   68 &  0.0 &  6.3 & 9.6\\ 
EIS 0541$-$2440 & 05 & 41 & 32.4 & $-$24 & 40 & 36.8 & 0.5 &  67.1 &   45 &  4.1 &  3.2 & 3.9\\ 
EIS 0541$-$2300 & 05 & 41 & 52.6 & $-$23 &  0 & 25.9 & 0.2 &  32.6 &   37 &  5.7 &  4.5 &  $-$\\ 
EIS 0543$-$2322 & 05 & 43 &  8.9 & $-$23 & 22 & 18.6 & 0.3 &  31.6 &   15 &  4.1 &  3.9 & $-$\\ 
EIS 0543$-$2324 & 05 & 43 & 20.3 & $-$23 & 24 &  3.5 & 0.2 &  24.2 &   30 &  4.2 &  5.5 & $-$\\ 
EIS 0543$-$2429 & 05 & 43 & 29.9 & $-$24 & 29 & 38.7 & 0.2 &  30.4 &   36 &  5.3 &  4.7 & 5.8\\ 
EIS 0543$-$2322 & 05 & 43 & 37.6 & $-$23 & 22 & 22.9 & 0.7 &  81.5 &   48 &  4.1 &  $-$ & $-$\\ 
EIS 0543$-$2326 & 05 & 43 & 45.2 & $-$23 & 26 & 13.1 & 0.6 &  68.9 &   42 &  2.9 &  4.3 & 3.3\\ 
EIS 0543$-$2322 & 05 & 43 & 58.6 & $-$23 & 22 & 22.1 & 0.2 &  30.0 &    9 &  5.3 &  $-$ & $-$\\ 
EIS 0946$-$2029 & 09 & 46 & 12.8 & $-$20 & 29 & 49.9 & 0.2 &  72.8 &   44 &  9.1 &  8.6 & 8.6\\ 
EIS 0946$-$2133 & 09 & 46 & 31.1 & $-$21 & 33 & 10.8 & 0.2 &  33.8 &   21 &  4.5 &  4.9 & 4.4\\ 
EIS 0946$-$2007 & 09 & 46 & 38.1 & $-$20 &  7 & 16.0 & 0.5 & 131.1 &   55 &  7.8 &  $-$ &  3.1\\ 
EIS 0947$-$2120 & 09 & 47 &  6.9 & $-$21 & 20 & 55.7 & 0.2 &  43.7 &   50 &  5.8 &  5.5 & 6.2\\ 
EIS 0947$-$2059 & 09 & 47 & 14.5 & $-$20 & 59 & 51.0 & 0.3 &  32.5 &   26 &  3.5 &  3.5 & $-$\\ 
EIS 0947$-$2030 & 09 & 47 & 47.3 & $-$20 & 30 &  4.0 & 1.0 & 133.3 &   23 &  3.1 &  3.2 & 3.6\\ 
EIS 0947$-$2109 & 09 & 47 & 58.6 & $-$21 &  9 &  6.2 & 0.2 &  27.2 &   14 &  4.0 &  5.4 &  5.2\\ 
EIS 0948$-$2044 & 09 & 48 &  8.8 & $-$20 & 44 & 31.2 & 0.2 &  43.4 &   32 &  5.5 &  5.9 & 6.0\\ 
EIS 0948$-$2113 & 09 & 48 & 22.1 & $-$21 & 13 & 55.0 & 0.3 &  41.2 &   53 &  4.7 &  $-$&  $-$\\ 
EIS 0948$-$2004 & 09 & 48 & 42.5 & $-$20 &  4 & 12.7 & 1.2 & 257.8 &  $-$99 &  4.1 &  2.6 &  5.3\\ 
EIS 0949$-$2121 & 09 & 49 &  1.7 & $-$21 & 21 & 47.2 & 0.4 &  66.7 &   25 &  5.3 &  3.3 & 3.5\\ 
\hline \hline
\end{tabular}
\end{table*}

\newpage
\begin{table*}
\addtocounter{table}{-1}
\caption{Continued.}
\label{tab:cluster1}
\begin{tabular}{lr@{\extracolsep{1mm}}r@{\extracolsep{1mm}}rr@{\extracolsep{1mm}}r@{\extracolsep{1mm}}rrrrrrr}
\hline \hline
Cluster name & \multicolumn{3}{c}{$\alpha$ (J2000)} &
\multicolumn{3}{c}{$\delta$ (J2000)} & $z$ & $\Lambda_{cl}$ & $ N_R $ & $\sigma_{even}$ & $\sigma_{odd}$ & $\sigma_{paired}$ \\ 
\hline
EIS 0949$-$2139 & 09 & 49 &  7.5 & $-$21 & 39 & 44.6 & 0.6 & 116.7 &   80 &  5.1 &  $-$ & $-$\\ 
EIS 0949$-$2101 & 09 & 49 & 22.1 & $-$21 &  1 & 47.1 & 0.3 &  37.7 &   57 &  3.6 &  3.9 &  4.0\\ 
EIS 0949$-$2101 & 09 & 49 & 22.9 & $-$21 &  1 & 50.1 & 0.3 &  29.3 &   25 &  3.5 &  3.2 &  3.7\\ 
EIS 0949$-$2118 & 09 & 49 & 28.0 & $-$21 & 18 & 47.6 & 0.4 &  40.7 &   45 &  3.3 &  3.0  & $-$\\ 
EIS 0949$-$2153 & 09 & 49 & 38.0 & $-$21 & 53 & 44.6 & 0.9 & 129.3 &   32 &  3.6 &  3.3  & 3.4\\ 
EIS 0949$-$2046 & 09 & 49 & 51.5 & $-$20 & 46 & 40.6 & 0.2 &  48.8 &   24 &  5.2 &  5.3  & 5.4\\ 
EIS 0950$-$2129 & 09 & 50 & 16.0 & $-$21 & 29 & 14.0 & 0.4 &  43.1 &   32 &  3.4 &  3.5  & $-$\\ 
EIS 0950$-$2113 & 09 & 50 & 23.6 & $-$21 & 13 & 54.3 & 0.4 &  57.3 &   43 &  4.8 &  4.0  & $-$\\ 
EIS 0950$-$2154 & 09 & 50 & 47.9 & $-$21 & 54 & 39.0 & 0.5 &  69.4 &   33 &  3.5 &  4.2  & 4.0\\ 
EIS 0950$-$2133 & 09 & 50 & 48.0 & $-$21 & 33 & 13.1 & 0.2 &  31.5 &   10 &  4.1 &  3.5  & 4.4\\ 
EIS 0950$-$2138 & 09 & 50 & 53.6 & $-$21 & 38 &  0.7 & 0.5 &  54.2 &   27 &  $-$ &  4.1  & $-$\\ 
EIS 0951$-$2052 & 09 & 51 &  8.3 & $-$20 & 52 & 23.6 & 0.2 &  31.9 &   16 &  3.4 &  3.2  & $-$\\ 
EIS 0951$-$2026 & 09 & 51 & 28.9 & $-$20 & 26 & 33.0 & 0.2 &  44.2 &   13 &  4.3 &  4.9  & 4.7\\ 
EIS 0951$-$2145 & 09 & 51 & 46.4 & $-$21 & 45 & 27.1 & 0.2 &  62.0 &   42 &  8.1 &  7.7  & 7.7\\ 
EIS 0952$-$2005 & 09 & 52 & 19.3 & $-$20 &  5 &  4.7 & 0.4 &  65.1 &   37 &  4.5 &  4.4 &  4.5\\ 
EIS 0952$-$2114 & 09 & 52 & 29.5 & $-$21 & 14 & 30.6 & 0.3 &  38.7 &   35 &  2.6 &  4.2  & $-$\\ 
EIS 0952$-$2032 & 09 & 52 & 32.6 & $-$20 & 32 & 40.0 & 0.3 & 106.9 &  156 &  9.0 &  9.7  & 9.3\\ 
EIS 0952$-$2121 & 09 & 52 & 36.1 & $-$21 & 21 & 59.6 & 0.4 &  72.8 &   31 &  4.9 &  5.8  & 5.0\\ 
EIS 0952$-$2138 & 09 & 52 & 37.3 & $-$21 & 38 & 25.6 & 0.4 &  52.1 &   33 &  4.2 &  4.4  & 4.6\\ 
EIS 0952$-$2048 & 09 & 52 & 41.1 & $-$20 & 48 & 25.9 & 0.4 &  55.5 &   37 &  3.0 &  4.5  & $-$\\ 
EIS 0952$-$2150 & 09 & 52 & 46.8 & $-$21 & 50 & 15.1 & 0.2 &  33.8 &   28 &  2.8 &  4.4  & $-$\\ 
EIS 0952$-$2103 & 09 & 52 & 47.5 & $-$21 &  3 &  3.1 & 0.2 &  33.4 &   11 &  3.6 &  3.2 &  $-$\\ 
EIS 0953$-$2054 & 09 & 53 &  6.0 & $-$20 & 54 & 17.3 & 0.2 &  50.2 &    6 &  5.4 &  4.9  & $-$\\ 
EIS 0953$-$2024 & 09 & 53 & 15.3 & $-$20 & 24 & 33.5 & 0.3 &  63.8 &   33 &  6.2 &  6.9  & 6.6\\ 
EIS 0953$-$2048 & 09 & 53 & 34.6 & $-$20 & 48 & 51.7 & 0.3 &  46.2 &   49 &  6.2 &  $-$  & 5.5\\ 
EIS 0953$-$2154 & 09 & 53 & 35.8 & $-$21 & 54 & 44.7 & 0.3 &  54.5 &   20 &  4.3 &  5.7  & $-$\\ 
EIS 0953$-$2032 & 09 & 53 & 49.7 & $-$20 & 32 & 40.0 & 0.3 &  35.1 &   25 &  3.5 &  3.8  &4.0\\ 
EIS 0953$-$2114 & 09 & 53 & 52.7 & $-$21 & 14 & 46.1 & 1.0 & 172.1 &   59 &  3.1 &  3.4 & 3.6\\ 
EIS 0953$-$2017 & 09 & 53 & 55.5 & $-$20 & 17 & 20.3 & 0.2 &  36.1 &   14 &  5.2 &  4.6 & 5.6\\ 
EIS 0954$-$2111 & 09 & 54 & 15.3 & $-$21 & 11 & 42.1 & 0.5 &  73.2 &   66 &  5.1 &  $-$ & 5.8\\ 
EIS 0954$-$2051 & 09 & 54 & 19.5 & $-$20 & 51 & 57.1 & 0.5 &  62.2 &   34 &  3.9 &  3.2 & 3.6\\ 
EIS 0954$-$2113 & 09 & 54 & 57.5 & $-$21 & 13 & 11.2 & 0.5 &  96.5 &   82 &  6.2 &  4.3 & 5.2\\ 
EIS 0955$-$2123 & 09 & 55 &  2.3 & $-$21 & 23 &  6.7 & 0.2 &  53.3 &   26 &  6.9 &  7.5 & 7.5\\ 
EIS 0955$-$2037 & 09 & 55 & 16.9 & $-$20 & 37 &  4.1 & 0.2 &  37.1 &   37 &  5.3 &  4.1 & 4.7\\ 
EIS 0955$-$2144 & 09 & 55 & 19.2 & $-$21 & 44 & 34.5 & 0.6 &  85.0 &  107 &  4.3 &  4.3&  4.0\\ 
EIS 0955$-$2013 & 09 & 55 & 30.7 & $-$20 & 13 & 51.1 & 0.5 &  51.9 &   31 &  3.1 &  3.2 & 3.8\\ 
EIS 0956$-$2054 & 09 & 56 &  2.7 & $-$20 & 54 &  8.6 & 0.2 &  38.1 &   28 &  5.5 &  5.0 & 5.6\\ 
EIS 0956$-$2026 & 09 & 56 & 24.0 & $-$20 & 26 & 58.7 & 0.2 &  31.8 &   19 &  4.6 &  4.6 & 5.6\\ 
EIS 0956$-$2101 & 09 & 56 & 24.9 & $-$21 &  1 & 11.7 & 0.4 &  53.4 &   22 &  4.3 &  4.0 &  4.5\\ 
EIS 0956$-$2059 & 09 & 56 & 25.2 & $-$20 & 59 & 49.8 & 0.3 &  39.7 &   37 &  4.7 &  4.5 & 4.8\\ 
EIS 0956$-$2009 & 09 & 56 & 28.6 & $-$20 &  9 & 27.4 & 0.5 &  58.2 &   22 &  3.5 &  3.0 &  3.9\\ 
EIS 0956$-$2050 & 09 & 56 & 42.0 & $-$20 & 50 & 47.9 & 0.2 &  27.2 &   48 &  2.8 &  6.5 & $-$\\ 
EIS 0956$-$2044 & 09 & 56 & 56.9 & $-$20 & 44 & 17.7 & 0.5 &  96.0 &   61 &  3.1 &  6.0 & $-$\\ 
EIS 0956$-$2107 & 09 & 56 & 57.9 & $-$21 &  7 & 33.3 & 0.3 &  30.1 &   51 &  3.0 &  3.6 &  4.0\\ 
EIS 0957$-$2051 & 09 & 57 &  7.2 & $-$20 & 51 & 44.3 & 0.2 &  27.9 &   36 &  4.0 &  4.9 & $-$\\ 
EIS 0957$-$2143 & 09 & 57 & 14.5 & $-$21 & 43 & 27.5 & 0.2 &  51.2 &   23 &  6.0 &  7.2&  6.7\\ 
EIS 0957$-$2016 & 09 & 57 & 30.3 & $-$20 & 16 & 25.5 & 0.6 &  96.7 &   47 &  5.6 &  $-$&  $-$\\ 
\hline \hline
\end{tabular}
\end{table*}

\begin{table*}
\caption{$3\sigma$ cluster candidates for EIS patches C and D.}
\begin{tabular}{lr@{\extracolsep{1mm}}r@{\extracolsep{1mm}}rr@{\extracolsep{1mm}}r@{\extracolsep{1mm}}rrrrrrr}
\hline \hline
Cluster name & \multicolumn{3}{c}{$\alpha$ (J2000)} &
\multicolumn{3}{c}{$\delta$ (J2000)} & $z$ & $\Lambda_{cl}$ & $ N_R $ & $\sigma_{even}$ & $\sigma_{odd}$ & $\sigma_{paired}$ \\ 
\hline
EIS 0532$-$2303 & 05 & 32 & 28.9 & $-$23 &  3 & 23.1 & 1.1 & 191.5 &  $-$99 &  $-$ &  $-$ &  3.0\\ 
EIS 0533$-$2359 & 05 & 33 &  0.3 & $-$23 & 59 & 32.5 & 0.7 &  81.0 &   22 &  $-$ &  $-$ & 3.3\\ 
EIS 0533$-$2307 & 05 & 33 &  6.9 & $-$23 &  7 & 21.1 & 1.2 & 283.7 &  $-$99 &  $-$ &  $-$ &  3.5\\ 
EIS 0533$-$2317 & 05 & 33 & 24.0 & $-$23 & 17 & 52.3 & 0.7 &  92.1 &   35 &  $-$ &  $-$ & 4.5\\ 
EIS 0533$-$2417 & 05 & 33 & 24.8 & $-$24 & 17 & 58.1 & 0.4 &  33.0 &   26 &  $-$ &  $-$ & 3.2\\ 
EIS 0533$-$2411 & 05 & 33 & 26.9 & $-$24 & 11 & 50.3 & 0.5 &  47.7 &   24 &  $-$ &  $-$ & 3.3\\ 
EIS 0533$-$2412 & 05 & 33 & 40.3 & $-$24 & 12 & 43.8 & 1.1 & 299.2 &   72 &  $-$ &  $-$ & 3.3\\ 
EIS 0533$-$2441 & 05 & 33 & 41.5 & $-$24 & 41 & 51.8 & 0.7 &  77.2 &   86 &  3.0 &  2.8 & $-$\\ 
EIS 0533$-$2258 & 05 & 33 & 43.2 & $-$22 & 58 & 45.4 & 0.9 & 107.3 &   46 &  3.1 &  $-$ & $-$\\ 
EIS 0533$-$2418 & 05 & 33 & 46.0 & $-$24 & 18 & 52.1 & 0.5 &  70.3 &   66 &  $-$ &  $-$ & 5.2\\ 
EIS 0533$-$2355 & 05 & 33 & 54.4 & $-$23 & 55 & 33.1 & 0.8 &  86.9 &   84 &  $-$ &  $-$&  3.5\\ 
EIS 0534$-$2430 & 05 & 34 & 12.5 & $-$24 & 30 & 17.9 & 1.0 & 173.4 &  $-$99 &  3.0 &  2.9 & $-$\\ 
EIS 0534$-$2405 & 05 & 34 & 55.3 & $-$24 &  5 & 20.6 & 0.7 &  81.1 &   54 &  $-$ &  $-$ &  3.3\\ 
EIS 0535$-$2402 & 05 & 35 &  7.4 & $-$24 &  2 & 28.4 & 0.4 &  46.3 &   45 &  2.6 &  3.9 &  $-$\\ 
EIS 0535$-$2344 & 05 & 35 & 11.0 & $-$23 & 44 &  9.6 & 1.2 & 282.6 &   12 &  $-$ &  $-$ & 3.5\\ 
EIS 0535$-$2335 & 05 & 35 & 33.9 & $-$23 & 35 &  6.2 & 0.3 &  31.5 &   18 &  3.8 &  $-$ & $-$\\ 
EIS 0535$-$2413 & 05 & 35 & 39.1 & $-$24 & 13 & 17.0 & 0.5 &  47.8 &   37 &  $-$ &  3.1 & $-$\\ 
EIS 0535$-$2302 & 05 & 35 & 46.4 & $-$23 &  2 &  9.2 & 0.3 &  31.6 &   30 &  2.7 &  3.8 &  $-$\\ 
EIS 0536$-$2334 & 05 & 36 & 40.4 & $-$23 & 34 & 39.3 & 0.5 &  42.5 &   29 &  3.3 &  $-$ & $-$\\ 
EIS 0537$-$2331 & 05 & 37 & 16.8 & $-$23 & 31 &  1.0 & 0.2 &  44.6 &   31 &  $-$ &  $-$& 10.2\\ 
EIS 0537$-$2354 & 05 & 37 & 17.9 & $-$23 & 54 & 46.2 & 0.8 &  83.3 &   29 &  3.7 &  2.6 & 3.4\\ 
EIS 0537$-$2444 & 05 & 37 & 28.0 & $-$24 & 44 & 18.8 & 0.3 &  22.6 &    6 &  3.0 &  2.6 & $-$\\ 
EIS 0538$-$2334 & 05 & 38 &  0.1 & $-$23 & 34 & 33.0 & 1.1 & 173.9 &  $-$99 &  $-$ &  3.0&  $-$\\ 
EIS 0538$-$2405 & 05 & 38 &  5.1 & $-$24 &  5 &  6.8 & 0.8 & 100.2 &   44 &  3.0 &  2.8 &  3.2\\ 
EIS 0538$-$2345 & 05 & 38 & 12.0 & $-$23 & 45 &  6.2 & 0.9 &  97.5 &   43 &  3.3 &  2.8 & $-$\\ 
EIS 0538$-$2331 & 05 & 38 & 48.0 & $-$23 & 31 & 41.1 & 0.5 &  37.4 &   16 &  2.8 &  3.1&  $-$\\ 
EIS 0538$-$2304 & 05 & 38 & 49.0 & $-$23 &  4 & 10.2 & 0.7 &  61.8 &   37 &  2.9 &  3.0 &  3.3\\ 
EIS 0538$-$2404 & 05 & 38 & 51.0 & $-$24 &  4 & 53.0 & 1.1 & 255.6 &  $-$99 &  $-$ &  3.5 &  $-$\\ 
EIS 0539$-$2341 & 05 & 39 &  $-$ & $-$23 & 41 & 31.7 & 0.6 &  47.0 &   38 &  $-$ &  $-$ & 3.2\\ 
EIS 0539$-$2313 & 05 & 39 &  1.8 & $-$23 & 13 & 56.0 & 0.4 &  31.3 &   17 &  $-$ &  $-$ & 3.7\\ 
EIS 0539$-$2348 & 05 & 39 &  5.8 & $-$23 & 48 &  5.8 & 0.2 &  64.1 &   23 &  $-$ &  $-$& 14.6\\ 
EIS 0540$-$2308 & 05 & 40 &  7.6 & $-$23 &  8 & 10.4 & 0.7 &  87.5 &   56 &  $-$ &  $-$ &  3.7\\ 
EIS 0540$-$2418 & 05 & 40 &  8.5 & $-$24 & 18 & 19.3 & 0.6 &  83.8 &   40 &  $-$ &  $-$&  4.9\\ 
EIS 0540$-$2310 & 05 & 40 & 11.4 & $-$23 & 10 & 48.3 & 0.4 &  30.8 &   37 &  $-$ &  $-$ & 3.4\\ 
EIS 0540$-$2343 & 05 & 40 & 18.5 & $-$23 & 43 & 13.1 & 0.6 &  64.0 &   12 &  $-$ &  3.4 & $-$\\ 
EIS 0540$-$2315 & 05 & 40 & 20.8 & $-$23 & 15 & 11.5 & 0.7 &  79.2 &   32 &  $-$ &  $-$ & 3.3\\ 
EIS 0540$-$2405 & 05 & 40 & 29.4 & $-$24 &  5 & 50.1 & 0.2 &  27.5 &   17 &  $-$ &  $-$ &  5.0\\ 
EIS 0540$-$2309 & 05 & 40 & 57.1 & $-$23 &  9 &  4.8 & 0.6 &  60.9 &   36 &  $-$ &  $-$ &  3.3\\ 
EIS 0541$-$2437 & 05 & 41 & 15.2 & $-$24 & 37 & 32.0 & 0.5 &  48.0 &   63 &  2.5 &  3.0 & $-$\\ 
EIS 0541$-$2432 & 05 & 41 & 16.8 & $-$24 & 32 & 32.1 & 0.4 &  44.4 &   40 &  3.7 &  2.9 & 3.1\\ 
EIS 0541$-$2316 & 05 & 41 & 39.0 & $-$23 & 16 &  6.2 & 0.2 &  38.3 &   25 &  $-$ &  $-$ & 7.3\\ 
EIS 0541$-$2305 & 05 & 41 & 48.6 & $-$23 &  5 & 35.5 & 0.8 &  89.6 &   52 &  $-$ &  $-$ &  3.1\\ 
EIS 0541$-$2400 & 05 & 41 & 55.8 & $-$24 &  0 & 36.7 & 1.2 & 489.5 &  $-$99 &  $-$ &  $-$ &  4.7\\ 
EIS 0542$-$2300 & 05 & 42 & 34.3 & $-$23 &  0 & 20.1 & 0.3 &  35.2 &   21 &  $-$ &  $-$ &  4.8\\ 
EIS 0542$-$2436 & 05 & 42 & 53.6 & $-$24 & 36 & 57.7 & 0.2 &  34.1 &   13 &  $-$ &  $-$ & 5.8\\ 
EIS 0543$-$2359 & 05 & 43 & 29.6 & $-$23 & 59 & 51.7 & 1.1 & 243.1 &  $-$99 &  3.2 &  2.9&  3.9\\ 
EIS 0543$-$2421 & 05 & 43 & 35.8 & $-$24 & 21 & 20.3 & 1.0 & 167.7 &   22 &  $-$ &  3.1&  $-$\\ 
EIS 0544$-$2426 & 05 & 44 &  2.8 & $-$24 & 26 & 51.0 & 0.2 &  33.3 &   28 &  $-$ &  $-$ & 5.6\\ 
EIS 0945$-$2005 & 09 & 45 & 58.0 & $-$20 &  5 & 38.5 & 0.9 & 124.2 &   62 &  2.7 &  3.7 &  3.5\\ 
EIS 0946$-$2103 & 09 & 46 &  2.0 & $-$21 &  3 & 18.9 & 1.0 & 152.5 &   45 &  3.9 &  $-$ &  $-$\\ 
EIS 0946$-$2053 & 09 & 46 &  5.2 & $-$20 & 53 & 41.5 & 0.6 &  66.5 &   49 &  3.3 &  2.9 & 3.5\\ 
EIS 0946$-$2023 & 09 & 46 & 45.4 & $-$20 & 23 & 54.8 & 0.3 &  36.7 &   30 &  3.9 &  $-$ & $-$\\ 
EIS 0947$-$2044 & 09 & 47 & 17.4 & $-$20 & 44 &  5.0 & 0.9 & 103.6 &   39 &  2.7 &  3.4 & 3.3\\ 
EIS 0947$-$2025 & 09 & 47 & 17.6 & $-$20 & 25 & 41.8 & 0.9 & 113.9 &   60 &  $-$ &  3.1 & $-$\\ 
EIS 0947$-$2057 & 09 & 47 & 34.9 & $-$20 & 57 & 24.3 & 0.4 &  43.9 &   29 &  2.7 &  3.5 & $-$\\ 
EIS 0948$-$2123 & 09 & 48 & 17.5 & $-$21 & 23 & 33.3 & 0.4 &  40.5 &   26 &  2.9 &  3.0&  3.1\\ 
\hline \hline
\end{tabular}
\end{table*}

\newpage
\begin{table*}
\addtocounter{table}{-1}
\caption{Continued.}
\label{tab:cluster2}
\begin{tabular}{lr@{\extracolsep{1mm}}r@{\extracolsep{1mm}}rr@{\extracolsep{1mm}}r@{\extracolsep{1mm}}rrrrrrr}
\hline \hline
Cluster name & \multicolumn{3}{c}{$\alpha$ (J2000)} &
\multicolumn{3}{c}{$\delta$ (J2000)} & $z$ & $\Lambda_{cl}$ & $ N_R $ & $\sigma_{even}$ & $\sigma_{odd}$ & $\sigma_{paired}$ \\ 
\hline
EIS 0948$-$2129 & 09 & 48 & 24.1 & $-$21 & 29 & 14.8 & 0.7 &  80.7 &   36 &  3.2 &  $-$&  $-$\\ 
EIS 0948$-$2151 & 09 & 48 & 46.8 & $-$21 & 51 & 34.7 & 1.0 & 203.7 &  $-$99 &  3.4 &  $-$&  3.1\\ 
EIS 0949$-$2147 & 09 & 49 &  0.0 & $-$21 & 47 & 10.7 & 1.1 & 239.3 &   76 &  $-$ &  $-$ & 3.4\\ 
EIS 0949$-$2058 & 09 & 49 & 32.6 & $-$20 & 58 & 29.5 & 0.5 &  58.2 &   63 &  2.9 &  3.3 & 3.2\\ 
EIS 0949$-$2117 & 09 & 49 & 51.6 & $-$21 & 17 & 24.1 & 0.5 &  63.1 &   88 &  $-$ &  3.9 & $-$\\ 
EIS 0950$-$2103 & 09 & 50 &  8.0 & $-$21 &  3 & 40.7 & 0.2 &  23.8 &    2 &  3.3 &  2.6 &  $-$\\ 
EIS 0950$-$2018 & 09 & 50 & 14.4 & $-$20 & 18 & 39.2 & 1.1 & 170.6 &  $-$99 &  $-$ &  3.5 & $-$\\ 
EIS 0950$-$2108 & 09 & 50 & 19.9 & $-$21 &  8 & 12.0 & 1.1 & 192.4 &   40 &  $-$ &  $-$ &  3.1\\ 
EIS 0950$-$2123 & 09 & 50 & 23.5 & $-$21 & 23 & 58.2 & 0.6 &  60.2 &   39 &  $-$ &  $-$ & 3.2\\ 
EIS 0951$-$2146 & 09 & 51 &  3.8 & $-$21 & 46 &  8.9 & 0.6 &  53.4 &   20 &  3.3 &  2.7 & 3.3\\ 
EIS 0951$-$2102 & 09 & 51 & 30.8 & $-$21 &  2 & 54.4 & 0.4 &  49.4 &   35 &  3.4 &  2.7 &  3.0\\ 
EIS 0951$-$2046 & 09 & 51 & 31.8 & $-$20 & 46 & 56.6 & 0.5 &  65.1 &   25 &  3.5 &  2.7 & 3.3\\ 
EIS 0951$-$2016 & 09 & 51 & 38.1 & $-$20 & 16 &  2.8 & 1.3 & 321.6 &  $-$99 &  $-$ &  $-$ & 3.7\\ 
EIS 0952$-$2047 & 09 & 52 & 32.5 & $-$20 & 47 & 34.1 & 0.7 &  79.6 &   49 &  $-$ &  $-$ & 3.1\\ 
EIS 0952$-$2012 & 09 & 52 & 36.3 & $-$20 & 12 & 57.2 & 1.1 & 168.7 &  $-$99 &  $-$ &  3.4 & $-$\\ 
EIS 0952$-$2144 & 09 & 52 & 48.7 & $-$21 & 44 & 32.6 & 0.2 &  36.1 &   41 &  $-$ &  $-$ & 4.7\\ 
EIS 0952$-$2009 & 09 & 52 & 49.0 & $-$20 &  9 & 27.2 & 0.9 &  93.0 &   17 &  $-$ &  $-$ &  3.2\\ 
EIS 0953$-$2052 & 09 & 53 &  4.4 & $-$20 & 52 & 48.5 & 0.2 &  38.9 &   19 &  $-$ &  $-$ & 4.2\\ 
EIS 0953$-$2105 & 09 & 53 &  6.3 & $-$21 &  5 & 29.2 & 1.0 & 186.0 &  $-$99 &  3.8 &  $-$ &  $-$\\ 
EIS 0953$-$2145 & 09 & 53 &  9.0 & $-$21 & 45 & 49.7 & 0.3 &  31.2 &   14 &  $-$ &  $-$ & 3.4\\ 
EIS 0953$-$2041 & 09 & 53 & 51.5 & $-$20 & 41 & 52.1 & 0.4 &  41.5 &   49 &  $-$ &  3.6 & $-$\\ 
EIS 0954$-$2120 & 09 & 54 &  1.2 & $-$21 & 20 &  3.2 & 0.9 & 110.4 &   42 &  2.5 &  3.1 & $-$\\ 
EIS 0954$-$2011 & 09 & 54 & 32.8 & $-$20 & 11 & 13.3 & 0.8 &  91.6 &   21 &  $-$ &  3.6 & $-$\\ 
EIS 0954$-$2023 & 09 & 54 & 47.5 & $-$20 & 23 & 55.2 & 1.1 & 217.9 &   52 &  $-$ &  $-$ & 3.3\\ 
EIS 0955$-$2033 & 09 & 55 & 15.5 & $-$20 & 33 & 59.8 & 0.4 &  42.4 &   31 &  $-$ &  $-$ & 3.8\\ 
EIS 0955$-$2137 & 09 & 55 & 19.1 & $-$21 & 37 & 59.7 & 0.5 &  43.6 &   36 &  $-$ &  $-$ & 3.0\\ 
EIS 0955$-$2113 & 09 & 55 & 32.3 & $-$21 & 13 & 55.3 & 0.6 &  68.2 &   67 &  $-$ &  3.2 & $-$\\ 
EIS 0955$-$2008 & 09 & 55 & 36.3 & $-$20 &  8 & 26.2 & 1.0 & 153.3 &   26 &  3.2 &  2.6 &  $-$\\ 
EIS 0955$-$2109 & 09 & 55 & 51.1 & $-$21 &  9 & 57.4 & 0.7 &  71.5 &   67 &  $-$ &  $-$ &  3.1\\ 
EIS 0956$-$2024 & 09 & 56 & 11.9 & $-$20 & 24 & 21.2 & 1.2 & 252.7 &  $-$99 &  3.0 &  $-$ & $-$\\ 
EIS 0956$-$2053 & 09 & 56 & 29.0 & $-$20 & 53 & 41.6 & 0.4 &  40.3 &   25 &  3.5 &  $-$ & 3.4\\ 
EIS 0956$-$2154 & 09 & 56 & 53.6 & $-$21 & 54 & 36.5 & 0.4 &  35.5 &   15 &  $-$ &  $-$ & 3.3\\ 
EIS 0957$-$2028 & 09 & 57 &  0.6 & $-$20 & 28 & 42.6 & 0.8 &  85.7 &   40 &  3.2 &  $-$ & $-$\\ 
EIS 0957$-$2104 & 09 & 57 &  1.0 & $-$21 &  4 & 37.6 & 1.1 & 188.9 &   25 &  3.1 &  $-$ &  $-$\\ 
EIS 0957$-$2005 & 09 & 57 &  9.6 & $-$20 &  5 & 29.4 & 1.2 & 256.6 &  $-$99 &  3.2 &  $-$ &  3.7\\ 
EIS 0957$-$2038 & 09 & 57 & 11.9 & $-$20 & 38 & 46.6 & 0.5 &  58.8 &    9 &  3.6 &  $-$&  $-$\\ 
EIS 0957$-$2044 & 09 & 57 & 15.8 & $-$20 & 44 & 54.4 & 0.3 &  36.1 &   17 &  3.9 &  $-$ & $-$\\ 
EIS 0957$-$2013 & 09 & 57 & 26.2 & $-$20 & 13 & 21.5 & 0.6 &  62.3 &   28 &  $-$ &  $-$ & 3.5\\ 
EIS 0957$-$2132 & 09 & 57 & 31.0 & $-$21 & 32 & 40.9 & 0.3 &  38.3 &    6 &  $-$ &  $-$ & 4.2\\ 
\hline \hline
\end{tabular}
\end{table*}


\begin{thebibliography}{}

\bibitem{benoist} Benoist, C., et al. 1998, submitted to A\&A;
(Paper~VI)
\bibitem{nonino} Nonino, M., et al. 1998, submitted to A\&A;
astro-ph/9803336 (Paper~I)
\bibitem{olsen} Olsen, L.F., et al. 1998a, submitted to A\&A;
astro-ph/9803338 (Paper~II)
\bibitem{olsenb} Olsen, L.F., et al. 1998b, submitted to A\&A;
astro-ph/9807156 (Paper~V)
\bibitem{postman} Postman, M., Lubin, L.M., Gunn, J.E., Oke, J.B.,
Hoessel, J.G., Schneider, D.P., Christensen, J.A. 1996, AJ, 111, 615
\bibitem{prandoni} Prandoni, I., et al. 1998, submitted to A\&A;
astro-ph/9807153 (Paper~III)
\bibitem{renzini} Renzini, A. \& da Costa, L. N. 1997, Messenger 87, 23
\end{thebibliography}
\end{document}